\documentstyle[12pt,psfig]{article}
\let\section=\subsection     \let\subsection=\subsubsection                
\def\be{\begin{eqnarray}}
\def\ee{\end{eqnarray}}
\newcommand {\p} {{\vec p}}

\newcommand {\k} {{\vec k}}

\begin{document}
\begin{center}
   {\large \bf NEUTRON-PHONON INTERACTION IN 
               NEUTRON STAR CRUSTS}\\[5mm]
   A.~SEDRAKIAN \\[5mm]
{\small \it  Center for Radiophysics and Space Research,\\
Cornell University, Ithaca 14853, NY\\[8mm] }
\end{center}

\begin{abstract}\noindent
The phonon spectrum of Coulomb lattice in
neutron star crusts above the neutron drip density 
is affected by the interaction with the ambient 
neutron Fermi-liquid. For the values of the neutron-phonon 
coupling constant in the range  $0.1 \le \lambda \le 1$ 
an appreciable renormalization of the phonon spectrum occurs which 
can lead to a  lattice instability manifested in an exponential growth of 
the density fluctuations. The  BCS phonon exchange mechanism of 
superconductivity leads to neutron pairing with a 
gap in the neutron excitation spectrum comparable to 
that due to the direct nuclear interaction.

\end{abstract}

\section{Introduction}

Below the melting temperature $T_m \simeq {Z^2\, e^2}/{100\, r_i}\sim 
10^9-10^{10} $ K, where $r_i$ is the average interion spacing, 
$Z$ is the ion charge, the ionic component of plasma in neutron 
star crusts is arranged in a Coulomb lattice. 
The screening of the electrostatic potential of the lattice by 
electrons is ineffective since the ratio 
$\lambda_D/r_i <1$, where $\lambda_D$ is 
the Debye screening  radius, and,  therefore, 
electrons are distributed almost uniformly. 
Above the neutron drip density $4 \times 10^{13}$ g cm$^{-3}$ the 
intervening space between the clusters is filled by a neutron fluid which 
goes over to a superfluid state below the critical 
temperature of the order $T_c\sim 10^9$ K;
(for a review of the early work see, e.g.,  ref. \cite{ST}; recent progress
is  summarized in ref. \cite{PR}). Although initial studies of 
the collective effects
in the crusts were focused either on the properties of the highly compressed 
solid matter or the unbound neutron fluid at subnuclear 
densities (with an exception 
of the equation of state where
phase equilibrium conditions are imposed) this separation is 
justified only when the interaction between these components 
of the crusts is weak;  this is not,  however, always the 
case. The purpose of this work is to continue the discussion 
of the interaction between the neutrons in the continuum and  
the excitations of the crustal lattice, i.e. phonons, set up 
in an earlier work \cite{AS}.

\section{The Phonon Spectrum}
The self-consistent coupled fermion-phonon problem is shown in terms of Feynman 
diagrams in Fig. 1 and 3. 
In this section we shall give the finite temperature solution of the Dyson 
equation for phonons which are coupled to fermions of arbitrary relativism;
in the present case the neutrons are only mildly
relativistic (if at all) while the electrons are ultrarelativistic. 
The problem can be solved 
for an arbitrary coupling strength since the perturbation series converge rapidly
with respect to another parameter - ratio of the phonon 
to the fermion energies: as a result one may consider only to 
the lowest order term in the integral equation 
for the vertex (Migdal theorem). 
The sequence in which the Dyson equations for the 
\begin{figure}
\begin{center}
\mbox{\psfig{figure=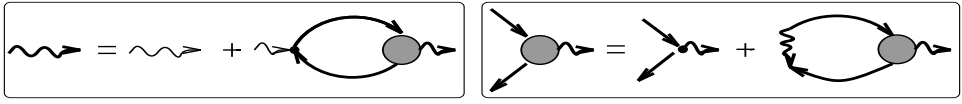,height=.5in,width=5.in,angle=-90}}
\end{center}
\caption[] 
{\footnotesize{The phonon Dyson equation and the vertex equation 
for coupled system of fermions (solid lines) and  phonon (wavy lines). The 
fermion Dyson equation is given by the block in Fig. 3.
The thick lines correspond to the full propagators, while the thin
lines to the free ones. }}
\label{fig2}
\end{figure}
fermions and phonons should be solved is fixed by the fact that 
the phonons affect
only a narrow range of energies in the fermion propagator (of the order 
the Debye frequency $\omega_D$), which implies that 
the polarization function can be evaluated with free fermion  propagators:
\be 
\gamma^0G^{0}(\p, i\omega_{\nu}) = \frac{\Lambda_+({\vec p})}
{i\omega_{\nu}-(\epsilon_p-\mu)}; \quad
\Lambda_+({\vec p}) =\frac{1}{2} \left\{1- 
\frac{\vec\alpha\cdot\vec p}{\epsilon_p}
      + \frac{\gamma_0\, m}{\epsilon_p}\right\},
\ee
where  $\epsilon(p)=\sqrt{p^2+m^2}$, 
${\vec\alpha} = \gamma_0\, {\vec \gamma}$ and 
we have kept only the positive energy states
from the outset; (to excite an excitation at the top of the Fermi sea and 
an anti-particle at rest one needs to overcome 
an energy barrier of the order of the 
Fermi energy $\epsilon_F$, however $\omega \le \omega_D\ll \epsilon_F$).
The Matsubara frequencies assume discrete 
odd integer values  $\omega_{\nu} = (2 \nu  + 1)/
\beta$  where $\nu = 0, \pm 1,\pm 2, \dots $ and $\beta$ is the 
inverse temperature;
the other notations have their usual meaning. 
We find the  retarded polarization function depicted in Fig. 1 
by performing the frequency summation over the product of 
the fermion propagators and  an analytical continuation 
to the real axis:
\be\label{11}
\Pi^{(R)} (\k, \omega) \! & = & \! {\cal M}_k^2
\int\!\!\frac{d^3p}{(2\pi)^3}\frac{f(\epsilon_{p+k}-\mu)
-f(\epsilon_p-\mu)}{\omega -\epsilon_{p+k}+\epsilon_{p}+i\delta}
 \left\{1-\frac{\p\cdot (\p +\k)}{\epsilon_p\epsilon_{p+k}}
+\frac{m^2}{\epsilon_p\epsilon_{p+k}}\right\}, \nonumber 
\ee
where ${\cal M}_k$ is the 
fermion-phonon coupling matrix element
and $f(x) = [1+ \exp(\beta x)]^{-1}$ is the Fermi distribution function.
The scalar polarization function above corresponds to particle--hole 
excitation for arbitrary degree of relativism of the system and 
finite-temperatures; for applications to the neutron subsystem one may take 
the non-relativistic limit. Fig. 2
shows the real part (more precisely $1+ {\rm Re} \Pi({\bf k},\omega)$)
and the imaginary part of the polarization function as a function of 
the momentum transfer for $\beta^{-1} = 0$ and a typical ratio $c_s/v_F = 0.1$, 
where $c_s$ and $v_F$ are the sound and Fermi velocities, respectively. 
The first quantity gives the renormalization of the phonon frequencies 
$[\omega(k)/\omega_0]^2$ (where $\omega_0$ is the unperturbed phonon frequency)
which are treated in the Debye model.
The eigen modes of the system vanish first in the long-wave limit where, with 
increasing coupling constant, the curves cross zero. The disappearance of the 
real solutions to the dispersion relation indicates instability of the system 
with respect to the density fluctuations, as can be seen by inspecting 
the proper solutions which exhibit 
exponentially growing amplitude of phonon mode.

\begin{figure}
\begin{center}
\mbox{\psfig{figure=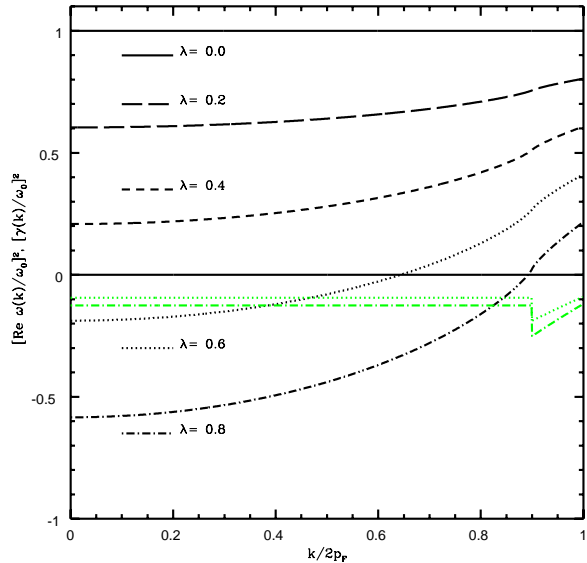,height=3.2in,width=3.2in,angle=0}}
\end{center}
\caption[] 
{\footnotesize{The renormalization of the phonon spectrum $\omega(k)$ and the
damping $\gamma (k)$
of the modes (light lines) as a function of the momentum transfer 
for several values of the neutron-phonon coupling constant. }}
\label{fig2}
\end{figure}

If this is the case, one may conclude that 
the starting lattice structure does not correspond to the true minimum of the 
energy of the system (in other words  is not the `true vacuum').
Available estimates of the neutron-phonon coupling constants are, however, 
unreliable  for two reasons: first, preliminary considerations
\cite{AS}, 
based on a fit to the elastic neutron-nucleus scattering 
cross-sections, does not include the medium effects 
(which most likely will reduce the effective cross-section)
and, in addition, the substantial imaginary part of the optical potential 
responsible for the absorptive processes should be included in the estimates.
Second, the compositions of the crustal matter due to different authors (see refs. 
in \cite{PR,AS}) imply coupling constants from  marginal up to of the order unity 
at $n_0/3$, where $n_0$ is the nuclear saturation density, and consequently, very 
different assessments about the role of the neutron-phonon interaction.

\begin{figure}
\begin{center}
\mbox{\psfig{figure=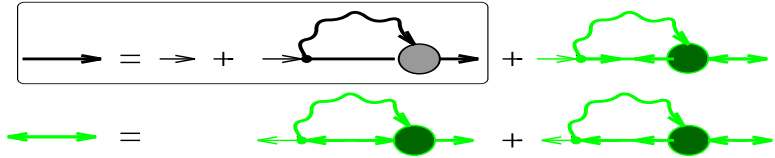,height=.8in,width=4.in,angle=-90}}
\end{center}
\caption[] 
{\footnotesize{The Dyson equations for the normal  and anomalous
fermion propagators. The contributions from the anomalous sector are shown by 
the light lines.}}
\label{fig2}
\end{figure}

\section{Phonon Exchange Interaction and Neutron Pairing}
The net two-body interaction between neutrons in the continuum comprises
the nuclear component (which is dominated by the attractive 
$^1S_0$ channel at densities of interest) and the component due to the 
phonon exchange. Since the latter interaction is  
attractive as well, the BCS pairing for some values of neutron-phonon
coupling constant might be influenced by the phonon-exchange mechanism.
The respective gap equations, which include the retardation of the 
effective interaction, emerge as a solution to the 
set of diagrams in Fig. 3. We find (for simplicity the $\beta^{-1}
=0$ limit is given below)
\be 
\Delta(\omega) z(\omega) &=& \int_0^{\infty}d\xi \!
~\left[{\cal K_+(\omega,\xi)}+{\cal W}\right]{\rm Re}
\frac{\Delta(\xi)}{\sqrt{\xi^2 - \Delta(\xi)^2}}
\nonumber \\ \label{GAP}
\omega\left[1-z(\omega)\right] &=& 
\int_0^{\infty}\! d\xi ~{\cal K_-(\omega,\xi)} {\rm Re}
\frac{\xi}{\sqrt{\xi^2 - \Delta(\xi)^2}}
\ee
where $z(\omega)$ is the wave function renormalization and 
the effective interactions via phonon exchange and direct nuclear
force  are defined respectively
\be 
{\cal K}_{\pm}(\omega, \xi) &=& 
 \frac{1}{ (2\pi)^2 v_F} \int_0^{2p_F}\!\! {\cal M}_k^2
\left[\frac{1}{\xi + \omega + \omega(k) + i\delta}\pm
\frac{1}{\xi-\omega+\omega(k)-i\delta} \right] k~dk\nonumber  \\
{\cal W} &=& \frac{1}{(2\pi)^2 v_F} 
\int_0^{2p_F}\!\vert V(k)\vert  k dk .
\ee
Here the nuclear interaction is given by the time-local interaction $V(k)$;
note that a pairing description based on the pion exchange model would 
allow for the effects of retardation, pion mode softening, condensation etc.
The formal structure of equations (\ref{GAP}) coincides with the Eliashberg
equations for metallic superconductor in the presence of  the repulsive 
Coulomb force between the electrons. 
For the values $\omega_D = 0.7$ MeV and $\lambda = 0.45$ (which correspond 
to the density $0.3 n_0$, for the composition of Arponen, see \cite{AS})
and ${\cal W} =0$, eqs. (\ref{GAP}) predict a value $\Delta = 0.03$ MeV, 
which is basically less, but comparable to the gap found 
neglecting the neutron-phonon coupling.

To conclude, if in a certain region of neutron star 
crusts the neutron-phonon
coupling is not negligible (i.e. $\lambda\le 1$), a number of 
novel effects emerge,  which are potentially important for the 
correct description of phenomena in neutron star 
crusts. The crucial problem is whether the
calculations of the coupling constants beyond the simple 
estimates would lead to significant 
values of the neutron-phonon coupling constant. 
In addition one may need a better understanding of the sources 
of the discrepancies between different models of neutron star 
crust composition and  a reassessment based on
the progress achieved in the many-body theory of nuclear matter in 
recent years.

\subsection*{Acknowledgment}
I would like to thank the organizers of Hirschegg '98 for their hospitality.
This work has been supported 
by the Max Kade Foundation (New York, NY).

\end{document}